# THE ORIGINS OF THE RESEARCH ON THE FOUNDATIONS OF QUANTUM MECHANICS (AND OTHER CRITICAL ACTIVITIES) IN ITALY DURING THE 1970s

(Revised version, October, 2016)

**Angelo Baracca\*, Silvio Bergia+ and Flavio Del Santo"**

\* University of Florence, Italy, baracca@fi.infn.it
+ University of Bologna, Italy, bergia@bo.infn.it
" University of Vienna, Austria, delsantoflavio@gmail.com

**Abstract**

We present a reconstruction of the studies on the Foundations of Quantum Mechanics carried out in Italy at the turn of the 1960s. Actually, they preceded the revival of the interest of the American physicists towards the foundations of quantum mechanics around mid-1970s, recently reconstructed by David Kaiser in a book of 2011. An element common to both cases is the role played by the young generation, even though the respective motivations were quite different. In the US they reacted to research cuts after the war in Vietnam, and were inspired by the New Age mood. In Italy the dissatisfaction of the young generations was rooted in the student protests of 1968 and the subsequent labour and social fights, which challenged the role of scientists. The young generations of physicists searched for new scientific approaches and challenged their own scientific knowledge and role. The criticism to the foundations of quantum mechanics and the perspectives of submitting them to experimental tests were perceived as an innovative research field and this attitude was directly linked to the search for an innovative and radical approach in the history of science. All these initiatives gave rise to booming activity throughout the 1970s, contributing to influence the scientific attitude and the teaching approach.

**Keywords**

Foundations of quantum mechanics; Varenna school on foundations of quantum mechanics; social engagement of young Italian physicists; Franco Selleri's critical attitudes; history of physics.

## 1. Introduction: motivation and goals; a turning point in Italian scientific research

In the successful book *How the hippies saved physics*, published in 2011 (Kaiser 2011),[1] David Kaiser describes how in the second half of the 1970s, after a long period of almost total indifference, the physicists' interest towards the Foundations of Quantum Mechanics (FQM) experienced a revival in the US. Kaiser reconstructs in great detail and with an abundance of documents and witnesses the origin of this revival, which derived from the cuts to research funds, and the consequent identity crisis among the young generation of physicists; a crisis that particularly animated the *Fundamental Fysiks* Group (FFG) in Berkeley, California. Surprisingly, it appears that this revival was rooted in a totally unconventional atmosphere of the widespread *New Age* mood, the Oriental mysticism, even the use of psychedelic substances, with the goal of achieving psychokinesis, transmission of thought and superluminal communication.

In spite of these somewhat metaphysical goals, "The group's intense, unstructured brainstorming sessions planted seeds that would eventually flower into today's field of quantum information science"(Kaiser 2011, Introduction, p. xvi).

Our present preliminary research has been motivated by the lesser-known fact that in Italy these interests and this field of research, not only survived the postwar years, but in fact had a revival at the turn of the 1960s, well before its manifestation in the US. As the two eldest between us were amongst the protagonists of that revival, we deem it relevant to leave a direct testimony, and to attempt a reconstruction of these events, even if much less ambitious than Kaiser's. In particular, it seems relevant to us that, although this anxious endeavour for renewal was, in Italy as in the US, rooted in the uneasiness of the youngest generation of the physicists for the prevailing mood of physical research (characterised by the race towards high energies), the Italian physicists had nevertheless a more realistic and concrete physical attitude than their successive American counterparts. Their aim was no less ideological, but at the same time more ambitious and yet more restricted. In fact, the young Italian physicists caught hold of the recently introduced *Bell's inequality* aiming at the possibility that Quantum Mechanics (QM) could show limits of validity. This could have opened the possibility of a new physical framework, and legitimised their criticism to the prevailing scientific attitude. The downside of their critique of QM was that they did not look, contrarily to their successive FFG colleagues, for new implications of *quantum entanglement*, which stimulated the basic results of the "third quantum revolution", albeit in rather unphysical terms.

---

[1] The book won the *Davis Prize* from the History of Science Society.

It seems revealing in this respect that one of the pioneers and most influential representatives of the Italian line of research on the FQM, Franco Selleri (1936-2013), expressly cited by Kaiser (Kaiser 2011, pp: 40, 208, 219, 270), became sympathetic with the unphysical proposals of the FFG (Herbert's "FLASH" experiment : Herbert 1982; Kaiser 2011, pp. 210 ff.), instead interpreting it as a possibility of falsifying QM.

Gian Carlo Ghirardi – who was the Italian physicist who anticipated the new implications of QM (Kaiser 2011, pp. 208-215, 221-225, 233, 235, 270), actually focused later his attention to these interests (from the mid 1970s, therefore subsequent to the period that we will analyse), and had a less radical point of view, which objectively resulted to be more fruitful. This will be addressed in further detail in section 13.2.1, dedicated to a summary of the Italian research activities after this significant decade devoted to the FQM.

Nevertheless we deem that our although partial reconstruction of the new upswing of the research on the FQM in Italy between the end of the 1960s and the beginning of the 1970s can help to shed light on the ideological-political mood of the young generation of the Italian physicists at the time. It also aids to highlight their active involvement in the social-political events of that crucial period, as well as the aspect of their cultural and ideological engagement, which objectively renovated the public attitude towards science and its historical and social interpretation. In our opinion, the deep 'entanglement' of these stimuli, elaborations, and concrete initiatives in multiple fields was an important, and rather particular aspect of Italian history, which has affected Italian science, and possibly society, more deeply than it can be perceived by the prevalent sphere of the current Italian scientific community. In this respect, it seems to us that at present the future of scientific research, as well as that of the new generations of physicists, appears quite uncertain, with the apparently illogical and self-destructive choices of the Italian ruling class.

We must declare from the outset that our reconstruction is (probably inevitably) biased by the (although full of gaps) personal recollections of the two eldest authors (A. B. and S. B.). Furthermore, our recollection of archive documents is unfortunately far from complete. We hope that this contribution will stimulate more in depth research on this crucial period of Italian history.

The revival around the turn of 1970s of the interest towards FQM in Italy, and the first phase of its evolution, was rooted in a growing uneasiness of the new generation of the physicists towards the setting and the aims of the existing scientific theories but developed however mainly in institutional contexts, specifically in the environment of the Italian Society of Physics (SIF). In fact, the problem of the FQM had explicit repercussions on the decisions of the Steering Committee of SIF. In light of these facts, we have conducted careful research in the SIF archives in Bologna.

With the growth of independent research activities and autonomous forms of organization within the young generation of Italian physicists, approximately after 1972, their relationship with SIF was losing relevance. For this subsequent phase, no systematic recollection of documents and records exists. The correspondence and documents of important Italian physicists, now disappeared, like Marcello Cini and Franco Selleri, are conserved in archives which unfortunately are not adequately organized or accessible for specific research.

## 2. Italian precursors: academic interest towards the FQM during the 1950s and 1960s

In the immediate post-war years, it was specifically the engagement of Edoardo Amaldi (1908-1989) and his initiatives, mainly devoted to nuclear and particle physics, which breathed new life in the research into physics within Italy (Amaldi 1979; Amaldi *et al.* 1998).

Among the young post-war and post-Fermi Italian physicists, Piero Caldirola (1914-1984) was a scientist-humanist, with philosophical concerns,[2] who was active in Pavia and then professor of theoretical physics in Milan. He was entrusted with the editing of the entry "Quantum Mechanics" (*Quantistica, Meccanica*) for the *Enciclopedia Italiana* (Caldirola 1961), which resulted in a review of the proposals and problems of the interpretation of QM and especially of the measurement theory. The entry was published in various versions (also coauthored by Angelo Loinger) and also as a separate booklet, which circulated among the students of his course on theoretical physics and consequently personally stimulated one of the present authors, A. B., who studied and graduated in Milan.

Caldirola transmitted interest towards the FQM to his pupils, who, competent also in statistical physics, proposed in 1962 a mechanism to explain the reduction of the wave packet based on the stochastic irreversible behaviour of the macroscopic measuring device (Daneri *et al.* 1962).[3] This interpretation has fallen into disuse at present, but for several years was widely discussed (see e.g. Krips 2007; Auletta 2000, p. 260).

We must remark that Daneri, Loinger and Prosperi, along with the Caldirola's school, did not bear dissatisfaction towards QM and its validity, but only tried to resolve the residual open problems.

Also Bruno Ferretti, professor of theoretical physics at the University of Bologna, cultivated some interests towards the problems of the FQM, although it seems that he has not left any written evidence (as we further explain in section 3).

One could argue that in 1966 a paper by Bernard D'Espagnat on the theory of measurement in QM appeared in the Italian journal *Supplemento al Nuovo Cimento* (D'Espagnat 1966), but it did not raise attention among the Italian physicists, as far as we know (the interest toward D'Espagnat's works began later in Italy).

## 3. Early dissatisfaction of Franco Selleri for QM, and the seminars in Bologna (1969)

One must remark that Franco Selleri, when he was recently graduated in physics, gave important contributions to main stream theoretical elementary particle physics, working on the recently introduced "peripheral" one-pion model of proton-proton scattering (see e.g. Selleri 1961). However his critical attitude grew quite soon. We have no elements to reconstruct the birth of Selleri's uneasiness with QM, which should date back to 1962-1965, as he recalled in an interview[4] in 2003:

---

[2] One should remark that the Italian tradition of physics was double-sided**.** in the field of the epistemological and methodological concerns. On the one side, the main exponent, Enrico Fermi, was substantially indifferent to this kind of uneasiness: although he was the author of the introduction of QM in Italy, all his lecture notes are practically lacking of such kind of remarks. On the contrary, Ettore Majorana, although prematurely and mysteriously disappeared, exhibited deep concerns about this subject matter (see e.g, Maltese, 2010). On the other hand, Fermi's friend and colleague Enrico Persico authored one of the clearest lecture notes on wave mechanics, with an extremely effective didactic approach, that strongly influenced that by Caldirola.

[3] From Freire's comprehensive book on the "Quantum Dissidents" we got acquainted of a controversy between Eugen Wigner and Leon Rosenfeld, that incidentally involved also Daneri's *et al.* paper (Freire 2014, Sect. 4.4, pp. 156-161).

[4] Selleri, Franco. June 24, 2003. Interview by Olival Freire.

I decided I wanted to go to the United States. I applied to five different universities and I got five positive answers so that in 1962 I decided to go to Cornell. I was attracted by the presence there of the famous physicist Hans Albrecht Bethe. But then, slowly, I started to develop a critical attitude towards contemporary physics. I mean I had a very strong drive. I liked physics very, very much and I was very active, but it was soon evident that there were problems, fundamental problems in physics. Already in 1965 I could see them.

It was only after he came back from a period of international experiences[5] and eventually moved from the University of Bologna to the one of Bari,[6] presumably in 1969, that his active engagement in FQM started (see section 13.1.1).

An event providing a testimony of the interest of Italian physicists for FQM took place in Bologna at the Institute of Physics (now the Department of Physics and Astronomy) during the first half of 1967. The event consisted of a cycle of seminars directed by the Professor of Theoretical Physics, Bruno Ferretti. Each of the components of the Theoretical Group was assigned the reading of a paper or a subject.[7] We have not been able to find detailed documentation of it, and in particular of its possible promotion by some of the young researchers active in the Institute, Franco Selleri among them. In the quoted interview, he stated:

Already in Bologna before coming to Bari I started saying I did not believe anymore that quantum mechanics was correct. And the reactions were negative.

A biographer of Selleri reports that:

[his interests], initially addressed to the field of phenomenology of microphysics, since 1968 shifted […] towards the FMQ […] Such a shift of interests ripened […] during a stage in Sweden where he got in touch with the ideas of Bernard D'Espagnat thanks to his famous book *Conceptions de la physique contemporaine* (D'Espagnat 1965). (Nutricati 1998, p. 38).

Selleri's dissatisfaction towards QM become manifest in a short note on this subject which appeared in as early as 1969, in which he declared:

No physical phenomenon is known which disagrees with the predictions of quantum mechanics (Q.M.). Nevertheless, several physicists found it very hard to accept this theory as basically correct. […] Several hidden-variable theory have been proposed, starting in 1926. […] Furthermore an explicit model-theory with hidden variables has been constructed by BOHM and BUB. It is therefore clear that the soundness of such theories must be judged on their success in predicting new experimental facts. […] Even though the theories of hidden variables are not completely developed, an important shift of philosophical attitude can be noticed: particle and waves are now objectively existing entities. (Selleri 1969a: sent to the journal on May 10, published in the issue of 11 June)

He also sent the preprint to Louis de Broglie, who, "in his response dated 24[th] of February 1969, [...] manifests his great interest towards the research if the Italian physicist and the intent to intensify the contacts" (Nutricati 1998, p 38).

In the same year Selleri delivered a much more substantial and detailed series of lectures in Frascati between June and July on *Quantum Theory and Hidden Variables* (Selleri, 1969b), of which the contents are remarkable:

---

[5] Selleri was awarded a grant at the Theoretical Division of CERN (Geneva) in 1959-61, then "Collaborateur Etranger" in Saclay (France) in 1962-63 and Research Associate at the Cornell University (NY) in 1963-65). (See http://www.uniurb.it/Filosofia/sellericurr.htm)

[6] Our colleagues Augusto Garuccio and Luigi Romano are putting in order an archive of documents that Selleri left in Bari: we hope that they will find documents that can throw light mainly in these transition years, when he had loosened his relationship with his previous students in Bologna, and had not yet established his new relationships in Bari.

[7] By interviewing colleagues who took part in the initiative, we found that there was a general interest of the whole theoretical group towards the foundations of QM, and that Professor Ferretti moved systematically critical options. Baracca, in particular, was assigned a paper by J. M. Jauch and C. Piron, *Can hidden variables be excluded in quantum mechanics?* (Helv. Phys. Acta 36, 1963, 470-485), attempting to demonstrate the incompatibility of hidden variables within QM.

> […] the fact that one cannot assume a realistic point of view about the particles participating in our experiments is extremely unpleasant to many of us. You became a physicist to discover the 'secrets of Nature' and after many years you find that there was no Nature and, thus, nothing to discover. [...]
>
> We start from the following Realistic postulate: An elementary particle is always associated to a wave objectively existing. […] 'Objectively' is taken to mean: independently on all observers. […] The wave has to be thought of as a real entity in some kind of postulated medium. Thus wave and particle are reminiscent of a boat in a lake. Boat and wave are both objectively existing and are found to be associated, in the sense that you cannot find a boat without a wave; the opposite is, however, possible.

Since 1969, Selleri had therefore expressed a general dissatisfaction about the validity of quantum mechanics, becoming interested in *hidden variables theories* and trying to propose experiments with the aim of verifying them. He explicitly expressed a *realistic postulate*, according to which a particle is always associated with an objectively existing wave; however, in Selleri's opinion, this wave is not an observable, but a *hidden variable*.

## 4. The evolution of the general context, in society and in the world of scientific research

We deem it necessary at this point to digress on the evolution of the general situation in society and in the world of research in those years, as it deeply influenced the atmosphere as well as the choices of the younger generation of physicists.

On a global level, there were deep influences of the student protests (Berkeley protests of 1964, and the "French May" of 1968), and also due to the radicalization of the labour struggles in Italy ("Hot Autumn" of 1969), which created direct links between workers and students, and highlighted the struggles and demands of the world of technicians and researchers.

The criticism of the role of science in capitalistic society, associated with an increasing dissatisfaction towards the methods and the purposes of scientific research, stimulated a true search for alternative ideas and practices within the younger generation of researchers. The situation developed on multiple and interconnected planes.

The criticism of the FQM was conceived (probably in different measures for each subject) as a feature of the critical analysis of the ideological aspects of physics, as an aspect of its class character, and in general of the limits of science, as opposed to the dominant ideology of progress.

The connection between these different aspects really emerged on October 30$^{th}$, 1968 during the annual Congress of the Italian Physical Society (SIF) held in Rome. In the proceedings of the Society one can read that the assembly of the associates

> is suddenly interrupted after an exponent of the Student Movement asks for the floor and turmoils follow. The Steering Committee (SC, Consiglio di Presidenza) of SIF convenes an extraordinary meeting in order to decide if to suspend the National Congress. (SIF, Bollettino, n. 62 1968, p. 7)

## 5. Selleri's initiatives (1969-1970) for promoting the course in Varenna of 1970 (besides other political initiatives)

In 1969 Selleri took initiatives aimed at promoting a specific course on the FQM among the international summer schools held by SIF in Varenna. He had been elected in the Steering Committee (SC) of SIF at the end of 1968 (SIF, SC 1968, p. 48) and he kept the position until 1970, when he left for the United States (he had been offered a visiting professorship at the University of Nebraska).[8]

---

[8] Incidentally, we must make a comment about a statement made by the then President of SIF, Prof Giuliano Toraldo di Francia, who at the time was President of SIF, in an interview with Freire (Freire 2014, p. 200), according to which at the time there was a risk that the Society could split due to political tensions. We do not agree with this Toraldo's statement, which seems to us exaggerated. The tensions were certainly very strong, as we also shall discuss, yet in our opinion certain concerns ran, although in various degrees, through the whole Italian scientific community. Therefore we maintain that

In the meeting of the SC on the 15th of March 1969, taking advantage of the fact that one of the three courses during the Summer of 1970 was not confirmed, "Selleri proposes a course on the Foundations of Quantum Mechanics" (SIF, SC 1969a, p. 180-184).

The proposal was approved in the meeting of the SC of 19 April 1969, in which:

> The proposal of a course 'Foundations of QM', already considered in the previous meeting of the SC, is examined again. Cini informs that he has already read the book by D'Espagnat and has found it very interesting. After a wide discussion, the SC expresses a clearly favourable opinion to the above-mentioned course and decides to propose Prof. D'Espagnat as a director, highlighting of considering interesting a comparison of the opinions of Daneri, Prosperi, Loinger and Wigner, Bohm, De Broglie. As Secretary of the course is proposed Selleri. (SIF, SC 1969b, pp. 147-48).[9]

It must be noted from the outset that Marcello Cini (1923-2012) was an already influential, although relatively young, theoretical physicist: his clear left-wing political and ideological positions, and especially the connection between his criticism of science and his political analysis, were to play a fundamental role in the subsequent evolution of the ideas and attitudes of the younger generations, as we will explore in the following sections (Gagliasso *et al.* 2015).

Another event clearly shows the complex interweaving of ideological and strictly political initiatives that characterized the spirit of those years. In the meeting of the SC on the 11th of October, 1969 (SIF, SC 1969c), regarding the financing of the courses in Varenna for-1970, Selleri posed the question if the SC had the intention of renewing, as in previous years, the application of funds to NATO (North Atlantic Treaty Organization), amounting to 13 million Italian Lire. It seems evident that the problem was out in the open: in fact, it raised a discussion inside the SC. Although the majority was in favour of renouncing NATO's funds, in view of the fact that the deadline for the application expired on October 15th, before the partners meeting at the end of October held at the SIF Congress, the CP decided in the meanwhile to file the application for 1970. Selleri put on record his opinion that "one could have renounced to the NATO's financing for 1970" (SIF, SC, 1969c, p. 87). Also in the SIF partners assembly in the 55° Congress in Bari on October 30th the problem raised a lively and long discussion, at the end of which the refusal was approved (SIF, Bollettino 1970, n. 74, pp. 6-29, minutes of the partners meeting 1969).

It seems appropriate to remark, also for subsequent developments, how the meetings of SIF between the end of the 1960s and the early 1970s were the natural place in which several concerns inside the Italian scientific community emerged and were confronted. Therefore, even with obvious contradictions, the young generation of physicists had early opportunities for discussion, elaboration and growth.

Regarding the involvement of scientists in military research, a special mention deserves the deep sensation that was raised, especially in the young generations, by the analysis published by Marcello Cini in September 1969, in the occasion of the first landing on the Moon, *Il Satellite della Luna* (Cini 1969). In this article, for the first time Cini denounced the deep economic and military interests, disguised as scientific curiosity, behind the race, in the hands of the major international powers, to the extraterrestrial space.

## 6. The international Varenna school on the FQM of 1970, a melting pot of elaboration for the young Italian physicists

---

the introduction of innovative or controversial items, that we have reconstructed in this research, was, at least by part of an "enlightened" component of the physicists, not merely instrumental. In any case, we deem also that the co-optation in the Steering Committee of SIF of declared left-wing physicists like Marcello Cini, and younger ones like Franco Selleri, was a subtle strategy.

[9] An interesting information provided in Freire's book (Freire 2014, p. 159), is that Leon "Rosenfeld was invited to speak about the measurability of quantum fields and accepted. However, after reading the invitation letter and the list of speakers, [among which there was Wigner] he withdrew and sent Jørgen Kalckar, a younger physicist from the Niels Bohr Institute, in his place" (cfr. footnote 3).

The Varenna course on the FQM opened on the 29th of June 1970. It constituted the first official interchange between the Italian community of physicists concerned about the FQM and the wider international community. The course was, for the already experienced physicists, a sort of international validation, but for the younger ones, apart from being a wide-ranging immersion in these problems, it was also an occasion for intense contact and exchange, and also for productive common elaboration on the themes of science, its history and its role in connection with society (as we will explore in the next section). More than likely for Selleri it was both things at once.

The participants received a letter from the Director of the course, Bernard D'Espagnat, of which the content is worth mentioning (D'Espagnat 1971, pp. xiii – xiv):

> [...] theoretical physics rests on three legs: experience, mathematics and a workable set of general ideas. Some would like to cut this third leg away [...] of course they are quite wrong".

> [...] under superficial agreement on how to use the rules [...] we entertain real differences of opinion as to what these rules refer to".

> [...] Let me suggest to you the following agreement: that we should not take as goals the conversion of the heretic but rather a better understanding of his standpoint; that we should not suggest that we consider as a stupid fool anybody in the audience (lest the stupid fools should in the end appear clearly to be ourselves!); that we should try to cling to facts; and that nevertheless we should be prepared to hear without indignation very nonconformist views which have no *immediate* bearing on facts".

The teachers of the school were: J. Andrade e Silva, B. D'Espagnat, B.S. De Witt, J. Ehlers, A. Frenkel, K.-E. Hellwig, F. Herbut, M. Jauch, J. Kalckar, L. Kasday, G. Ludwig, H. Neumann, G.M. Prosperi, C. Piron, F. Selleri, A. Shimony, H. Stein, M. Vigičić, E. Wigner, M. Yanase, H.D. Zeh. However the proceedings contained also chapters from Louis de Broglie and David Bohm.

The topics tackled in the school were organised into three groups of fundamental problems of QM: (i) measurement and basic concepts, (ii) hidden variables and non-locality, (iii) interpretation and proposals. For the first topic, Prosperi, among the Italian lecturers, reproposed the thesis, already put forward in the previous paper of 1962 (Daneri *et al.* 1962), of the interpretation of the measurement problem based on the macroscopic nature of the apparatus. Although his viewpoint was never directed towards a critique of the validity of QM, it is nevertheless remarkable how topical his concern was on the limits of the application of quantum theory[10]. The only other Italian among the teachers was again Selleri, who presented in the section of interpretation and proposals his lecture "Realism and the wave-function of quantum mechanics". Here he insisted on his realist position and showed, using the Bell's inequalities applied to particle physics (his background field), that "the empirical evidence for wave functions of the second type[11] is not overwhelming". (D'Espagnat 1971, p. 401)

Freire rightly comments in his book (Freire 2014, pp. 207-210) that Wigner played a central role during the course, both for his ideas on QM and the process of measurement (the step of the "realisation" of its result by the observer dominated the debate), and for the political climate. Freire rightly recalls that the American aggression to Vietnam was the hottest political issue and Wigner was a convinced supporter of the politics of the United States.[12] Both these aspects animated the evening discussions among the young Italian participants, and the document they elaborated, that we discuss in the following section.

---

[10] He strongly criticized the lack of a formal distinction between the apparatus (macroscopic) and the quantum system (microscopic). In fact he stated: "We can put the *boundary* between observed system and apparatus or between apparatus and sense organs or between sense organs and nervous system, etc. [...] the transition [projection measurement] would be determined, in the last analysis, by the *abstract ego* becoming conscious of the result." (D'Espagnat, 1971, p. 103)

[11] At the time, the *entangled states* were usually referred as wave functions of second type or improper mixtures.

[12] From Freire's book A.B. is vividly reminded of the heated disputes raised by Wigner's initiative of a party to celebrate the National Day of US on July 4th (Freire 2014, pp. 210-211).

## 7. Active participation of the young concerned Italian physicists

For the young concerned generation of Italian physicists, who participated in this event in large numbers, the school was the first big occasion to meet, to bond and to discuss in depth and elaborate on concrete ideas; it was a true melting pot (Freire 2014 cites, in section 6.3.1, several witnesses that the Italian participants were the most politicized among the participants). Among the unsettled physicists who attended were, besides Selleri: Angelo Baracca, Vincenzo Capasso, Gianni De Franceschi, Carlo De Marzo, Donato Fortunato, Gianni Mattioli, Alessandro Pascolini, Marta Restignoli, Tito Tonietti (a mathematician), Livio Triolo, Luigi Solombrino. Capasso and Fortunato were young students of Selleri in Bari (see section 13.1.1).

They gathered every evening and developed deep and animated critical discussions using as starting points the themes presented in the lectures, such as the contradictions and paradoxes of QM, generalizing their perspectives to include the basic concepts of science and knowledge, their nature and function, their evolution and their meaning, and the social responsibility of scientists in capitalistic society. From these discussions they produced a collective 12-page document, that was typewritten, duplicated, and distributed on one of the last mornings at the entrance of the lecture room to all the teachers and participants: *Notes on the connection between science and society* (VV. AA. 1970).

The aim of the document was to raise awareness among the physicists towards the limits of the restricted conception of pure science (separation between physics, philosophy and society, and the presumed objectivity of science, as an absolute and ahistoric concept) and to point out the strong relation, too often omitted, between science and politics, economy and society (non–neutrality of science, criticism to "scientism" and submission of scientists to the needs of the leading class). These ideas, which at present could appear ingenuous or trivial, had actually at that time a breaking value in the scientific community and they were partially rooted in those of the student protests, but here can be noted an effort to further develop and solidify them, in particular in the aspects of physics and its applications.

As can be read in the collective document:

> It is instead extremely important to realize that science is certainly not neutral[...]. the structures of a scientific theory reproduce the categories of the culture of the dominant classes. […] we have experienced this fact even in the discussions of the school of physics "E. Fermi" on the FMQ: the poverty of the philosophical background pushes people towards an attempt to separate Physics from Philosophy with an evidently arbitrary and artificial procedure. […] We have considered to some extent the role of our consciousness in photon polarization experiments, but we have not considered our consciousness involvement in phenomena such as the explosion of the atomic bomb. […] The limitation of the individual consciousness to laboratory activity, disregarding any judgement of the social application of research, results in indiscriminate support for all applications of science (e.g., atomic bomb, chemical and bacteriological warfare). […] The scientist's incapacity to control the product of his research facilitates its cultural manipulations and the creation of consensus. […] This mystification is formalized in the powerful theory ("scientism") which assumes the intrinsic ability of science to solve all the human problems. […] Now we conclude that a pre-decision is strongly needed concerning the social structure in which men live and act. That is, a pre-decision on the historical and social role of the scientist, on his responsibility, on the fact that no concept or activity is neutral." (VV. AA. 1970).

We think we are not overestimating the importance and the long-lasting consequences of the lively discussions that took place collaterally to the official teaching activities at the 1970 Varenna School on the FQM. It was the first opportunity in a considerable period of time that a significant number of the young concerned physicists had gathered together. It was an early seed that contributed to the spread of the need to develop more precise analysis and research.

## 8. The aftermath of the Varenna 1970 course, and the increasing interest in the FQM

Following the active interest in the course, some of the young participants started thinking of devoting themselves to research activity on the FQM in their future practice (apart from Selleri, who was already active; while Prosperi, being older, was not part of the concerned current). As documented by Freire 2014 (p. 212), "when d'Espagnat was editing the proceedings, he was afraid of the influence of the political context on the editing process", since according to Franco Selleri "a trend would have appeared in [the Italian Physical] society whose goal would be to intimately associate scientific activities with activities from a different order." The answer by the President of SIF, Toraldo di Francia (1916 – 2011, was an influential physicist, whose philosophical interests were to emerge soon), reported by Freire (p. 212), reflects the whole discussion that led to the organization of the course, that we have discussed in section 5:

> The directors of the Varenna's courses were, *always*, the final judges of what should be included in their proceedings and your case is not an exception. By the way, it is true that there is in our society the trend of not occupying itself *only* with technical issues, and this trend is more and more strong and well founded. However, surely, by foundations of quantum mechanics we understand the foundations of quantum mechanics, and this is the title of the proceedings to be published.

In fact, the subsequent years saw a blossoming of research activities on the FQM by part of the young Italian physicists, and the school helped network them, and connect them with the international environment, but promoted also collaborations between the more experienced participants (Freire 2014, p. 206).

One of the present authors (A.B.) felt the need to both critically recollect the basic ideas of the school and to share them at his re-entry from Varenna, delivering in Autumn of 1970 a series of lectures at the Institute of Physics of Florence University, and writing detailed lecture notes (Baracca 1970): Selleri's lectures in Frascati were cited among the references, and one out of the four chapters of the lectures was devoted in great detail to the theory of Daneri, Loinger and Prosperi.

Regarding Selleri, he really was a precursor in Italy (since Daneri, Loinger and Prosperi did not question the validity of QM), and created a school of fellows in his new position at the University of Bari. In 1970 they published a dense review article devoted to a general fine-tuning of the problem of hidden variables in QM, the experimental verification of Bell's inequality, and the concept of an objectively existing wave function (Capasso *et al.* 1970). For the following developments of Selleri's group on this subject see section 13.1.1.

*Note: a significant coincidence*

In a paper on the history of the quantum physics controversy, Olival Freire remarked a significant temporal coincidence of two events that had no direct relation between them: the school of Varenna on the FMQ, and the creation of the Journal *Foundations of Physics* (Freire 2003). In fact, both reflected concerns and dissatisfaction that were increasingly spreading in those years in the scientific community.

> "The scientific journal Foundations of Physics appeared in 1970 with the aim of being the vehicle for debates in the field designated by its title, and above all, theoretical debates related to quantum physics. [...] Its Editorial Board comprised physicists [...]on opposite sides in quantum disputes [...] David Bohm and Louis de Broglie, former causal supporters, side by side with V. A. Fock [...] near the complementarity interpretation. [...] the philosopher Karl Popper and the physicist Eugene Wigner [...] in the first volume, 16 out of the 18 papers dealt with quantum themes. [...] the analysis of "Bell's Inequalities" did not deserve attention, the opposite happened in Varenna"

> > "It's noteworthy that in spite of the differences between the Foundations of Physics and the Varenna's course, both needed to face the same task in order to justify their existence: to argue against the instrumentalistic view of science. The decade after the creation of Foundations of Physics and […] the Varenna course have confirmed how opportune they were."

We could add that it was the same *Foundations of Physics* which was then to publish in 1982 the controversial paper by Herbert on the "FLASH" experiment (see sections 1 and 13.2.1) on superluminal communication (Herbert 1982; see in Kaiser 2011, pp. 210 ff.)

## 9. The genesis and development of the proposal of the Varenna school on History of Physics (HP), and the Study Day of June 1970 on science and society

As we have anticipated, the interest for the bursting potential of the criticism to the FQM was only one of the themes the young generation of Italian physicists was concerned about, and in fact it intertwined with other far sighted initiatives on multiple levels. This context made it possible to place the proposal to organise a course in Varenna on the history of physics for the summer of 1972. In fact, on the $7^{th}$ of February 1970, during the meeting of the SC of SIF, Selleri acknowledged the requirement of "a further analysis on the concept of research" (*SIF, SC,1970a, p.* 182-84), which was previously expressed during the already mentioned Partners meeting of the society at the $55^{o}$ annual congress in Bari, in October 1969 (see section 5). It seems that even the highest positions inside SIF (and therefore some of the most important teaching and research positions at a national level) became aware of these issues, as in the same session the President of SIF Toraldo di Francia declared that he

> had thought to a course which could have as totally indicative title 'Philosophy and History of Science'. Such a course if well organized could represent something completely new for Varenna. The emphasis should be put on the social implications of the study and research in Physics. […] Two coordinators should also be appointed: one a physicist and one a historian." (*SIF, SC 1970a, p.* 183-84)

The following meeting of the Steering Committee on the $2^{nd}$ of March 1970 returned to this issue, postponing further decisions (SIF, SC 1970b, pp. 2-4), while Selleri presented a document proposing a meeting on the role of scientific research in the university.

The latter proposal was discussed in the subsequent meeting of the SIF's SC of March $26^{th}$, 1970, where the decision was taken to organise a *Study Day on Science in Capitalistic Society (SIF, SC, 1970c, p. 23-25),* which indeed took place on the $13^{th}$-$14^{th}$ of June 1970 (SIF, Bollettino, n. 76 1970) in Florence, of which the proceedings were published by the editing house De Donato (SIF, 1971). This meeting was divided into three main topics with each of them introduced by two exponents of different, more or less radical, points of view, and/or disciplines: scientific research and teaching (Ettore Casari, and Giunio Luzzatto), myth and reality of science as a source of welfare (Marcello Cini, and Siro Lombardini), and the cultural and social role of science (Silvio Bergia, and Giorgio Salvini).[13] There was a wide participation of the young physicists and an active and lively debate in each session.

Concerning the project of a Varenna course on the History of Physics, in the meeting of the SC of SIF on April $18^{th}$ 1970 (SIF, SC, 1970d, p. 54-57) the president Toraldo di Francia announced that he had found a real interest in the subject, and he met with Prof. Jona-Lasinio (a physicist-mathematician) and Prof. Rossi Monti (a philosopher, see below) with the purpose of organising this course in the summer of 1971. He even reported a possible detailed program for the course, in which the names of Thomas Kuhn, Karl R. Popper and Noam Chomsky appeared, associated to such topics as "authoritarianism in science" and "science and political power in the last two decades". Further discussion and the proposal of the name of a possible director of the course were postponed, and they took place in a subsequent meeting of the SC of the $31^{st}$ May 1970 *(SIF, SC 1970e, p.* 2-4).

---

[13] Ettore Casari (born in 1933) has been an eminent philosopher of mathematics and logic of the school of L. Geymonat (see section 12). Giunio Luzzatto (born in 1935) has been a professor of mathematical analysis. Siro Lombardini (1924 – 2013) was an economist. Giorgio Salvini (1920 – 2015) was an influential physicist and a politician.

However, in the meeting of July 18[th], 1970, the President informed the SC that difficulties arose during the organisation of the course for the following year (SIF, SC 1970f, p. 113). In the successive meeting on the 9[th] of September 1970 it was finally decided to postpone it to 1972 (SIF, SC 1970f, p. 140-41). These difficulties resonated also in the partners meeting of SIF in the 56º Congress held in Venice (SIF, Bollettino, n. 84 1971, p. 5 ff.). In fact in the following meeting of the SC on the 17th April of 1971, Toraldo announced the decision taken in the latter Partners meeting, and proposed to resume communication with Jona-Lasinio and Paolo Rossi (SIF, SC, 1971a, p. 66-67). In the meeting of the SC of May 8[th] 1971, appears Jona's suggestion of Prof. Charles Weiner as a director of the course (SIF, SC 1971b, p. 125-26), whose official acceptance was communicated in the meeting on the 25[th] of May 1971 (SIF, SC 1971c, p. 137).

It is worth mentioning, in the meanwhile, a self-organised initiative of the young dissatisfied physicists which was the origin of the following activities in the field of history of physics. A first self-organized national gathering took place in Florence (organized by one of the present authors, A. B.), probably in 1971, in which the idea of developing a Marxist (social) history of science was discussed, aiming to search for the roots of the birth and the development of science (especially modern one) in connection with those of capitalistic society. The participants were inspired by Marx' historical materialistic approach, but were unsatisfied of the traditional (mostly humanistic) Marxist school of the history of science, which relied on a conception of fundamental "neutrality" of science from the social-economic context (see sections 12 and 14).

## 10. The critique to the approach and praxis in the research on elementary particle physics

As we have anticipated, the dissatisfaction of the young generation took various, intertwined, forms and expressions. It spanned from direct political engagement towards the investigation on the connections between the development of modern science and capitalism. Furthermore these young scientists started criticizing the foundations of physics, as well as the methodological attitude and the praxis of research activity in physics of elementary particles (from which the majority of the young Italian physicists came from). The first expression of this uneasiness was formulated, as far as we could reconstruct, again by Selleri, who, as we have recalled, had brought relevant contributions into this field, and at that time had already spent some years in research stages abroad (see footnote n. 5, section 3). In 1970 he published a note in the SIF Bulletin with the expressive title *The Aztec pyramid of the theoretical elementary particle physics* (Selleri 1970), which raised some discussions in the research environment. In this paper he described in colourful terms the approach to this field of research as

> […] the apocalyptic vision of the whole, always increasing, Aztec pyramid launched at uncontrolled gallop on the road to Nothing while the light of the Truth has already disappeared below the horizon (Selleri 1970, p. 16).

Selleri's note evidently caught an anxiety that was spreading, as he almost immediately received a reply from Elio Fabri (Fabri 1970, p. 8-12).

Even the two elder authors of this paper (A. B. and S. B.) were perceiving an increasing uneasiness in their research activity on phenomenological theoretical elementary particles physics, and were strongly stimulated to make them explicit. For this reason they presented two critical technical communications at the following SIF congress of October 26[th] 1971 in L'Aquila (Baracca and Bergia 1971; Baracca *et al.* 1971). The large participation and the long and lively discussion following these two communications was a signal that some uneasiness regarding the practice of research activity was spreading inside the community of the physicists. The content of the two communications was then developed in more detailed form in two internal notes of the INFN (National Institute of Nuclear Physics) Section of Bologna (Baracca and Bergia 1972; Baracca *et al.* 1972). In the first of the two, the authors gave some comments about the attitude of colleagues active in those years, and some of them had something to say concerning the way research that was carried out. Paolo Guidoni of the University of Rome, in particular, had remarked that "with such an attitude, the initial stage undergoes a random approach to problems, that allows a fast progress in the initial phase, but subsequently turns out to be negative in order to optimize information" (Guidoni 1970).

These notes were the backbone of what, some years later, was to become a book, with the significant title *La Spirale delle Alte Energie* (Baracca and Bergia 1976, *The High Energy Spiral*).

As a matter of fact, Selleri abandoned his activity in particle physics at the turn of the 1960s, and Baracca and Bergia did the same some years later.

## 11. Development and contents of the school of Varenna of 1972 on the history of physics

As already mentioned in section 9, despite its initial issues, the course *History of Twentieth Century Physics* eventually took place on July 31$^{st}$- August 12$^{th}$ 1972 (Weiner 1977). The teachers of the School were: E. Amaldi, H.B.G. Casimir, P.A.M. Dirac, L. Kowarsky, V.F. Weisskopf (physicists); J.L. Bromberg, Y. Elkana, J.L. Heilbron, G. Holton, M.J. Klein, C. Weiner (historians of physics); P. Rossi (historian of science); W. Goldstein, and M.J.Sherwin (social historians).

As stressed by Charles Weiner, the director of the School,

> Together [the historians and physicists in the school] take into account the philosophical and historical roots of the concepts and techniques of physics, the style of individuals and institutions, and the influences of social and political environments. […]
>
> Another important feature of the Varenna summer school was the intense and spirited discussion that engaged a large proportion of the faculty and students in the informal evening sessions. The school took place during the Vietnam War, and the lectures on the social and political history of physics gave rise to concern about the role of physics in contemporary history. These discussions culminated in a statement, drafted by some of the participants, condemning the war and the use of physics to prosecute it. The statement was approved by most of the school's participants. (Weiner 1977, p. xi)

The historical subjects discussed spanned from the beginnings of quantum theory to the development of atomic and solid state physics, quantum electrodynamics and several other modern subjects, as well as the scientific revolutions of the 17th and 20th century; the relationship between science and technology, science, politics and international affairs; specifically, the atomic bomb. Several of the younger participants had already met at the of 1970 on the FMQ. As it had happened on that occasion, also during the course of 1972, the young participants like Angelo Baracca, Silvio Bergia, Giovanni Ciccotti, Guido Cosenza, Salvo D'Agostino, Vittorio De Alfaro, Carlo De Marzo, Elisabetta Donini, Claudio Garola, Giulio Giorello, Sandro Petruccioli, Carlo Tarsitani and Tito Tonietti (including, in this case, foreign colleagues, like Jean-Marc Lévy-Leblond and Fritjof Capra, besides Marcello Cini, who was somewhat older and already an experimented scientist) had intense discussions during the evenings. In this case, however, they did not produce a written document, since the direction of the school only authorized an open evening discussion among all the participants on the social responsibility of science. One of the themes discussed was the involvement of the physicists in weapon development.[14] It is however worth mentioning a written one-page statement, against the involvement of scientists in the development of war technologies used in Vietnam, which was signed by 58 participants of the school and dated 12$^{th}$Augurs 1972 (Freire 2014, p. 218).[15]

---

[14] We thank Antonino Drago for sharing his reminiscences in a personal communication to the authors, 30th September 2015.

[15] We however did not succeed to reconstruct the genesis of such a document during the course, and even less its possible connection with the mentioned evening discussions. None of the Italian participants in the Course that we were able to contact, as well as the two elder authors of this paper, A.B. and S.B., can remember either the document, or how it originated, by whom, and in what circumstances. They were: Antonino Drago (communication to the authors, 2nd September 2016), Guido Cosenza (communication to the authors, 21st September 2016), Gian Carlo Ghirardi (communication to the authors, 30th September 2016), Arcangelo Rossi (communication to the authors, 19th September 2016), and Livio Triolo (communication to the authors, 26th September 2016). Even Jean Marc Lévy-Leblond, who partly reproduced the statement in his book "(Auto)-Critique de la Science" (Éditions du Seuil,

**12. The early concrete researches in the history of physics**

This new Varenna course on the HP, in 1972, gathered many of the young physicists who had already taken part in the previous course of 1970 on the FQM, besides new participants. Some of them had already participated in previous meetings, like the Study Day of 1970 in Florence and/or the self-organised meeting in Florence (see section 9), and were to constitute the backbone of a national collective of physicists and mathematicians engaged in this field of research. It is necessary to remark that at the beginning of these activities, the young scientists had no previous experience as historians.[16]

On this basis a concrete activity of research began and grew in the following years. The early papers were published in Italian in the journal of SIF, *Giornale di Fisica,* which remained therefore a sort of a test for the beginning of these activities.[17] These activities document the ample background underlying the interests and views of the young physicists of the time. As a matter of fact, the activity on history of science was to become one of the main ones for both Baracca and Bergia.

It is necessary to remark that in Italy there was already a thriving academic Marxist school of history and philosophy of science (generally not specifically on physics and composed by humanistic scientists) founded by the acknowledged authority of Ludovico Geymonat (1908-1991), and other researchers, such as Paolo Rossi Monti (1923-2012) and Lucio Colletti (1924-2001), all of them with a humanistic background.[18] The radical positions developed by the majority of the young Italian physicists soon diverged from those of Geymonat, and soon after also from positions that had directly inspired them, expressed by Rossi Monti in *Storia e Filosofia: Saggi sulla Storiografia Filosofica* (1969, *History and Philosophy: Essays on Philosophical Historiography*), and by Colletti in *Il Marxismo e Hegel* (1969, *Marxism and Hegel*). It is interesting to remark a singular coincidence, although the school of Geymonat had not been involved in the course of Varenna on the FQM (some scholar, such as Giulio Giorello, participated in the course of 1972), a component of this school published in 1972 a book that recollected the debate on QM in the USSR and the Soviet materialistic interpretation (Tagliagambe 1972).

The divergence, and the polemics that followed, centred on the adoption of the point of view of "dialectical materialism", introduced by Marx and particularly by Hegel, that matured and gained popularity especially after the publication of the book *Attualità del Materialismo Dialettico*, by Geymonat and his most direct collaborators (Bellone *et al.* 1974). This philosophy was denounced by the young scientists-historians as totally insufficient, and equivalent to a legitimization of science, since the relationship between science and capitalistic society was disregarded. We will come back to the subsequent developments in the section 14.

---

Paris) in 1973, did not give us any further information on the genesis of the document (communication to the authors, 12th September 2016). We are thankful for all their testimonies.

[16] An exception was the younger Arcangelo Rossi, who graduated in philosophy of science in Pisa at Scuola Normale, making his PhD studies in Rome with Vittorio Somenzi.

[17] For Baracca the first concrete research began – after a preparatory paper (Baracca and De Marzo 1972) – in 1973, in a research for a didactic degree thesis with Riccardo Rigatti (Baracca and Rigatti 1974.a), in which they related the origin of the physical concept of *work* and *power* to the development of the 18th century industrial revolution, retracing the technical scientific contributions of the English "engineer" John Smeaton (1724-1792) in the *Transactions of the Royal Society* of 1759. They then began the search for the roots of thermodynamics to the steam engine and Watt's invention of the separate condenser (Baracca and Rigatti 1974.b). A few years later, Bergia and Paola Fantazzini, inspired by the above mentioned papers, came back to Smeaton in order to show that he had played an important role in the attempts to derive the basic conservation laws describing two-bodies collisions (Bergia and Fantazzini 1981; Bergia and Fantazzini 1982).

[18] This Marxist school of thought had very influential exponents, in general historians, who set the ideological basis for the "Italian way to communism", many of them were leaders of the Italian Communist Party, such as for example Eugenio Garin (1909-2004), the mathematician Lucio Lombardo Radice (1916-1982), Ernesto Ragionieri (1926-1975), Giuseppe Vacca. The new political movements ideologically clashed with their Marxist ideology: this happened, in particular but not only, specifically in the field of science.

## 13. The growth of the research activity on the FQM in Italy during the 1970s

During this period we have examined, the activity on the FQM flourished. Selleri's school developed, Baracca and Bergia established two steadily tight groups respectively in Florence and Bologna, which grew involving dedicated students.

### 13.1. The line of research on Bell's inequality and the experimental tests of QM

*13.1.1. Further activity of Selleri's group.*

After Selleri's lecture in 1970 at the Varenna school on the FQM and his previously cited review article of the same year on hidden variables theories with Capasso and Fortunato (section 6), he apparently had a couple of years of adjustment concerning the Bell-type inequality and its possible tests (Selleri 1972; Capasso *et al.*1973). Subsequently, starting from 1976, in collaboration with his pupil Augusto Garuccio and other collaborators (in particular, since 1980, with Nicola Cufaro-Petroni), Selleri published a total of 20 papers between 1976 and 1980. Of which we explicitly refer only to the first ones (Augelli *et al.* 1976; Fortunato and Selleri 1976; Garuccio and Selleri 1976; Fortunato *et al.* 1977; a detailed analysis of the full research activity of Selleri and his collaborators is given in a book by Nutricati, 1998).[19]

*13.1.2. The early activities in Bologna (Bergia) and Florence (Baracca) on the FMQ*

The first paper from the Baracca-Bergia collaboration appeared in 1974 (Baracca *et al.* 1974). The publication attempted to systematise the observable differences between "proper" and "improper" mixtures (the terms that were used at that time for the entangled states), in view of the experimental tests that were starting. As we have remarked from the outset, what characterized the Italian currents on the FQM was the expectation that QM could be finally found limited in its physical validity. Starting from the 4$^{th}$ of June 1974, a two-day national conference was held in the research centre of Frascati, gathering the Italian researchers interested in this field (Baracca, 1974). Their following research activity in this field – occasionally extending their collaboration to Marta Restignoli (an elementary particle physicist of Rome, who also matured dissatisfaction for her field of research), and associating new fellows – continued until 1980 – extending and generalising this research (Baracca *et al.* 1974; Baracca *et al.* 1976; Baracca *et al.* 1977; Baracca *et al.* 1978a; Baracca *et al.* 1978b; Bergia *et al.* 1979; Bergia et al. 1980).

Baracca, prompted by the course of Varenna on the FQM, also spent a two-month period in 1974 with David Bohm (1917 – 1992) in his institute in London, with the explicit aim of establishing a fruitful exchange. This stage produced a common paper (Baracca *et al.* 1975). Roberto Livi proposed in his degree thesis, supervised by Baracca in Florence, a Bell type inequality for multivalued observables, and a possible experimental test in molecular physics (Livi 1977; 1978).[20]

---

[19] For a list of publications of Augusto Garuccio see http://beta.fisica.uniba.it/Portals/1/Documenti/CV/CV_Augusto_Garuccio.pdf, and of Cufaro-Petroni http://www.ba.infn.it/~cufaro/scientific.html, (all accessed October 15, 2016).

[20] During the 1970s many other ferments and engagements intertwined with scientific activity, which are worth to be briefly mentioned. In the first place, the environmental problem arose, since the pioneering book by Rachel Carson, *Silent Spring* (Houghton Mifflin Company, 1962), denoted as "a book that changed the world". In Italy, while Dario Paccino's *L'Imbroglio Ecologico* (The Ecological Swindle, Torino, Einaudi, 1972) expressed distrust about pure conservation of nature, as a further instrument of capitalistic classes, the Greens began to appear. Fritjof Capra's best-seller *The Tao of Physics* (1975, translated into Italian in 1982) had in Italy a controversial resonance, since it was largely applauded by the environmentalist, while it did not meet the favour of the Italian physicists.

Another interest and engagement arose, in the second half of the decade, with the growing protests against nuclear power, until then considered as an indisputably safe technology by the scientific milieus. A group of students in Florence got interested in the critical problems of nuclear power, in a completely autonomous way performed a comprehensive assessment of this technology, and published a pioneering book (Ciliberto *et al* 1977). They confronted criticism from the community of physicists, but were soon involved in the nascent Italian anti-nuclear movements as "popular experts", measuring themselves favourably with professional nuclear experts. When, at the turn of the 1970s, the group of students graduated, Baracca substantially

### 13.1.3 The Erice school on FQM of 1976

An international course on the FQM was held in Erice from April 18 to 23 with the title *"Thinkshop on Physics"*[21] and directed by John Bell and Bernard D'Espagnat. Also Selleri and his school actively participated to the event., but the proceedings were never published .

Due mentioning is the essential international resonance that this event have had in the following developments, as Freire noticed:

> […] the Erice 1976 meeting, which was seen by some physicists as the turning point in the acceptance that quantum nonlocality was indeed a new physical effect. (Freire 2006)

Also John Clauser,[22] who had already given important contributions to the FQM (Freedman and Clauser 1972) and at that time was part of the FFG remembered:

> the sociology of the conference was as interesting as was its physics. The quantum subculture finally had come out of the closet. ( Clauser 1992, p. 172)

Also remarkable is the fact that Anton Zeilinger, who was to become one of the founders of the Quantum Information field, heard for the first time of entanglement at the course in Erice:

> Zeilinger went to Erice unaware of entanglement but came back fascinated by the subject. (Freire and Pessoa 2015)

## 13.2. Interpretations and developments of QM

### 13.2.1. Ghirardi and collaborators

Gian Carlo Ghirardi, as mentioned in section 1, arrived later to these interests and had a less radical point of view: in fact he accepted QM as a starting point for further developments and thus he never questioned its validity. He, with some coworkers, devoted their attention to the Quantum Foundations of the Exponential Decay Law (Fonda *et al.* 1973), the Implications of the *Bohm-Aharonov hypothesis* (Ghirardi *et al.* 1976), the quantum nonseparability for systems composed of identical constituents (Ghirardi *et al.* 1977) and the stochastic interpretation of quantum mechanics (Ghirardi *et al.* 1978).

It was in 1979 that Ghirardi started playing an important role in a crucial international controversy on FQM. In fact, as already mentioned, the new concepts of quantum non-locality brought Nick Herbert – among the other members of the *Fundamental Fysiks* Group (see Section 1) – to propose experiments supposed to be able to transmit superluminal signals. The most effective proposals were the two experiments called "QUICK" (Kaiser 2012, pp. 206 – 209) and "FLASH" (Herbert 1982; Kaiser 2012, pp. 210 ff.), developed in the spring of 1979 and in January 1981 respectively. This line of research caught the interest of Selleri, who presented the "QUICK" experiment in a seminar in Udine in as early as 1979 (Selleri 1979). Ghirardi attended the seminar and immediately opposed the idea, publishing a formal paper at the end of the same year in which he claimed the theoretical impossibility of Herbert's proposed experiment (Ghirardi and Weber 1979). Consequently, Ghirardi was chosen as one of the referees of Herbert's paper by the journal *Foundation of Physics* (and afterwards also Tullio Weber, when Herbert submitted his second experiment "FLASH"). In his referee's report on the 22[nd] of April 1981, Ghirardi recommended a strong rejection of the paper (see. Kaiser 2012, p. 213). The motivation

---

inherited their experience and public engagement, prosecuting an enduring activity in the anti-nuclear movements. This whole story is reconstructed in detail in Baracca *et al.* 2016.

[21] John Bell wrote a report of the workshop, and the abstracts of the papers were published in "Progress in Scientific Culture" (Journal of the Ettore Majorana Centre, 1/4, 439–460, 1976). We owe this reference to Freire (2014, footnote 82 p.269).

[22] John F. Clauser, Alain Aspect and Anton Zeilinger have been awarded the prestigious Wolf Prize in Physics in 2010, for their fundamental conceptual and experimental contributions to the foundations of quantum physics, specifically for a series of tests of Bell's inequalities, or extensions thereof, using entangled quantum states.

was based on formal proof of a fundamental flaw in Herbert's experiment, i.e. the impossibility of copying an arbitrary unknown quantum state. In this result one can see the core of what later has became famous as the "no-cloning theorem",[23] an extremely important achievement of modern QM. As a matter of fact, driven by the stimulus of Herbert's proposal, William Wooters and Wojciech H. Zurek firstly formulated this theorem (Wooters and Zurek 1982), immediately followed by an independent derivation of Dennis Dieks (Dieks 1982). The work on FQM of Ghirardi and its collaborators, in particular Rimini and Weber, in the 1980s, received an important international resonance. They developed a new description of the collapse of the wave-function in QM (Ghirardi *et al.* 1985; Ghirardi *et al.* 1986), called the "GRW theory". This theory belongs to the class of the so called theories of "objective collapse", in which the wave-function collapse spontaneously occurs at a random instant, according to an average time-rate proportional to the number of particles of the system under study, providing thus a possible solution to the macroscopic-measurement paradox (Schrödinger's cat).

### *13.2.2. Further activities, and collaborations of the Group of Bari*

During the 1970s, Selleri started building a series of international collaborations (Augusto Garuccio spent one year in Moskow, from June 1974)[24] and in particular with the *Fondation Louis de Broglie* in Paris (Selleri was in contact, since 1978, with George Lochak, Pierre Claverie, Simon Diner, Olivier Costa de Beauregard and he met several times with De Broglie himself).[25] A long lasting collaboration was established, in 1978, with Jean Pierre Vigier (1920-2004) still in Paris (Cufaro-Petroni and Vigier 1979; Garuccio and Vigier 1980; Selleri and Vigier 1980). This fully involved Nicola Cufaro-Petroni (who spent 1978-79 in Paris with Vigier) and Augusto Garuccio (who worked in Paris with Vigier in 1980). In fact, between 1979 and 1985, they published with Vigier 23 and 13 papers, respectively.

Moreover, since 1979, Garuccio began collaborating with Vittorio Rapisarda of the University of Catania for the design of an experimental test of Bell's inequality (Garuccio and Rapisarda 1981; see for a detailed discussion Nutricati 1998, p. 113-114). During the preparation of the experiment, called "Foca-2" (Falciglia *et al.* 1982; Garuccio *et al.* 1982), Rapisarda suddenly died in a car accident on a visit to Bari in 1982, therefore Garuccio continued his experimental activity with the group of Catania. The interest towards the experimental tests of Bell's inequality in Catania had been anticipated in as early as 1974 by Salvatore Notarrigo[26] (1931-1998) and Diego Gutkowski (*Faraci et al. 1974*).

### *13.2.2.1. Relation between Selleri (and the Group of Bari) and Popper, in the 1980s*[27]

Of great interest is the relationship established between the *dissidents* of Quantum Mechanics and the philosopher Karl Popper, which involved the group of Bari. To this relationship Olival Freire briefly refers in his book (Freire 2014) in terms in our opinion partially incorrect, while he develops a deeper analysis in a previous paper in Portuguese (Freire 2004). Freire´s most recent and accessible sentence is the following:

> In the 1980s [Selleri] was responsible for building a bridge between critics of quantum mechanics and the philosopher Karl Popper, who was himself concerned with this physical theory, bringing the controversy in quantum mechanics to a wider audience. Selleri and Sexl's widely translated book, *Die Debatte um die Quantentheorie*, was part of this endeavor (Freire 2014, p. 213).

---

[23] The theorem states that an arbitrary unknown quantum state cannot be copied. His consequences are central for the modern developments of Quantum information and cryptography (see. e.g. a qualitative introduction in Kaiser 2011, chapter 9).

[24] Augusto Garuccio, personal communication to the authors, October 12, 2016.

[25] Nicola Cufaro.Petroni, personal communication to the authors, October 12, 2016.

[26] Salvatore Notarrigo: A Life in Science, thematic session, XXIII National Congress of the Italian Society for the History of Physics and Astronomy, 7-14 September 2004, https://indico.cern.ch/event/241987/page/7, accessed 15th October, 2016.

[27] We owe the development of this section to a suggestion of one of the referees, who pointed out that Freire mentions in his book (Freire 2014) the development of relations between Selleri and Popper in the 1980s.

In the first place, but of minor relevance, Selleri and Sexl´s book is actually authored by Selleri alone[28], and Popper is not even cited in it. Most important is that it was most probably Jean Pierre Vigier who, around 1980, first brought Popper once again into the active debate on FQM within the community of the physicists[29]. In fact Karl Popper had moved criticisms towards the standard interpretation of QM since the 1930s, when he also had intense discussions with several physicists, Einstein and Bohr among them (see e.g. Shield 2014). Due to a mistake he made in a proposed *Gedankenexperiment,* Popper become discouraged for some time about his contributions to the FQM, but after the 1950s he kept working on his own on possible realist interpretations of QM. Such an activity was widely illustrated by Popper in his book *Quantum Theory and the Schism in Physics* (Popper 1982). And it is at the very beginning of the 1980s as well, that Popper started an active collaboration with some of the physicists concerned with the new critics to QM. The first documented participation of Popper to a research with some physicists dates back to 1980-81, and this led to the publication of a paper co-authored by Popper himself, Vigier and Garuccio (Garuccio *et al*. 1981). The latter, a collaborator of Selleri, participated to the project with Popper only because he was at that time working with Vigier in Paris, and he did not even meet the philosopher in that period.[30] At that time there was no connection between Popper and Selleri's group in Bari. As it is clear from Selleri´s correspondence with Popper, cited in Freire´s paper (Freire 2004).

Selleri established a direct relationship with Popper only in 1983,[31] when he invited him to a conference in Bari, which was held in May of the same year. In his introductory talk at this conference, Popper presented "a new variant of the EPR experiment that can decide between a realistic interpretation and Copenhagen" (Popper, in Tarozzi and van der Merwe 1985, p. 3). The same experiment was actually already published (Popper 1982, pp. 27-30), but in Bari the author was directly addressing the community of the physicists concerned with the interpretations of QM, and explicitly asked them for an experimental realisation of his proposed experiment. In subsequent years a few papers were published either supporting or criticizing Popper's experiment (Freire 2004, p. 122), which was finally carried out in 1999, five years after Popper's death, and had quite a lot of resonance (for a review on the debate about Popper's experiment see Shield 2014).

Selleri´s relationship with Popper (not free from tensions, Freire 2004, p. 123; Nutricati 1998, p. 40) had a further development when Selleri in 1984 invited him to write an introduction to the Spanish translation of his already cited book (Freire 2004, p. 120; the Spanish translation of Selleri´s book is Selleri 2006).

### 13.2.3. Bergia's further research in the 1980s

Around the end of the 1980s Bergia and collaborators in Bologna studied the implications of the non-standard interpretation of QM based on the stochastic theory, originally proposed by E. Nelson,[32] along which the evolution of a quantum mechanical system can be described in terms of a stochastic process. They produced a couple of papers on the subject (Bergia *et al*. 1988; 1989) and, in the second one, the authors state that a good candidate to play the role of quantum probability could be that of metric fluctuation, discussed in detail in the first paper. This viewpoint does not seem to have led to a fully satisfactory alternative version of quantum mechanics. To quote one aspect, present in the concluding remarks of the second paper, "the hypothetical background medium of stochastic mechanics cannot be independent of the test particles"(Bergia *et al.* 1989). Furthermore Bergia and coworkers devoted some attention to the subjects h*idden variables* and experimental tests of QM (Bergia and Cannata 1982; Bergia *et al.* 1985).

---

[28] The Austrian physicist Roman Sexl was the editor of the series of books "Facetten der Physik", where the book written by Selleri was published.

[29] We are grateful to Agusto Garuccio and Nicola Cufaro Petroni, who shared their recollections in a joint communication to the authors, on October 11, 2016.

[30] Augusto Garuccio, personal communication to the authors, October 11, 2016.

[31] One the present authors, F.D.S., is grateful to Karl Milford, a pupil and collaborator of Popper, for the interesting discussion and for the selected literature, and to David Miller, Popper's assistant, for his kind suggestions on the use of Popper Archive in Klagenfurt (personal communication to the author, October 2, 2016).

[32] E. Nelson, *Dynamical theories of Brownian motion*, Phys. Rev. 150 (1966)

**14. The crucial and enduring development of the activities in the history and the critique of science**

The critical activity developed by the young generation of Italian physicists (and mathematicians) was the main, most fruitful and enduring inheritance left by those early years. During the 1970s these activities extended to several fields, well beyond the FQM, of which we have discussed the genesis and the initial steps. For what concerns the analysis of the subsequent developments of these different activities, they partly fall outside the purpose of the present reconstruction, although they brought deep and enduring changes to scientific mentality. Today the concept of what was referred to as "non-neutrality" of science, which was harshly contested by the majority of the scientific community, has probably entered into common mentality, although the distance between such a critical viewpoint and the scientific ideology has not been reduced. During the 1970s, for instance, courses on the History of Science, which did not exist in the past in Italian scientific faculties as it was considered to be a field reserved for philosophers, have been institutionalized in practically every degree course in Physics, although with different approaches (rather, it is today that this inheritance is becoming problematic with the drastic cuts to the development of the Italian university, which are penalising in the first place every perspective for young people in "collateral" research fields).

A collaboration which had profound significance and results was that of Marcello Cini, Angelo Baracca and other researchers, between 1974 and 1977, for the completely new series of the old–popular journal Sapere, entrusted by the editing house to the deeply politically engaged physician Giulio A. Maccacaro[33] (1924-1977). The editorial board developed a far-sighted and multifaceted research activity on the critique and the political and social implications of science. A true and wide sector was promoted for the diffusion of the analysis developed from the new current of the history of science. A big resonance (also at an international level) had the publication in 1976 (the same year of the already cited Baracca and Bergia's La Spirale delle Alte Energie) by Marcello Cini, Giovanni Ciccotti, Michelangelo De Maria and Giovanni Jona-Lasinio of the collective volume L'Ape e l'Architetto (Ciccotti et al. 1976, The Bee and the Architect), whose title explicitly recalled the famous comparison of Marx, underlining a genuine "Marxian", historical materialistic view. In contrast, the already mentioned Attualità del Materialismo Dialettico by Geymonat and his school was inspired by "dialectical materialism" (Section 12) and was perceived as a less radical critique to science. We will not continue to elaborate on the contributions of the thriving and very active school of Geymonat (particularly their most direct and most influential pupils Enrico Bellone, Giulio Giorello and Silvano Tagliagambe). In 1976 Baracca and Arcangelo Rossi developed an alternative "historical materialistic" for natural sciences, intended as an extension of Marx' analysis of economic science (Baracca and Rossi 1976).

In the same period, what we have described as the self-organized group of History of Physics (Section 9) had generated a relatively more compact group, whose components developed concrete initiatives, that had growing resonance and enduring consequences. The group active at the University of Lecce (Arcangelo Rossi, Elisabetta Donini and Tito Tonietti) organized on July 1$^{st}$ to 5$^{th}$ 1975, an extremely remarkable initiative (Donini *et al.* 1977), which gathered a very large number of the concerned physicists and mathematicians, whose aim was to deepen the critical analysis of science that had been developed in connection with the political and scientific engagement that the Italian scientific community had developed in this period (in connection with this, new courses of mathematical methods of physics were created in the subsequent years).

---

[33] Maccacaro was a physician, biologist and biometrist who devoted his work to the critique of science and supported the working class struggles. For an insight about his figure see e.g., Mara, 2010.

It is impossible, and inappropriate, to mention here in detail all the subsequent production in this field of history of science, therefore we will just quote the most immediate results and activities. Baracca actively collaborated with Roberto Livi and Stefano Ruffo (Baracca and Livi 1976; Baracca *et al.* 1979), extending their historic reconstruction to the period of the first and the second industrial revolution, up to the First World War. Bergia took part in the national project for the teaching of relativity in the secondary school and developed other research in the history of physics. Worth mentioning are the further activities of Arturo Russo, Michelangelo De Maria, Giovanni Battimelli, Elisabetta Donini, Arcangelo Rossi, and others. Baracca, who taught the course of statistical mechanics at the University of Florence (since 1968) under the stimulus of the student movement and after several successive re-elaborated versions, published an original textbook in which he integrated the rigorous scientific aspects with the critical-historical development of the discipline (Baracca 1980).[34]

With the growth of activity in the history and critics of science, in 1978, the core of the group of history of science developed the need for a more specific and less limiting instrument than *Sapere* for progressing in an adequate way, sharing their elaborations and gathering other collaborations which adopted a similar point of view. As a result they succeeded to promote the publication of the four-monthly journal *Testi e Contesti*, which was published for 9 issues, before closing in 1983 due to economic difficulties (see Rossi 2010).

Finally, we arrive to the peak of the activity promoted and undertaken by the core of the group of history of science, when it organised a very ambitious initiative in 1980. It was an international conference, organised with exceptional effort in two venues in Florence and in Rome (since the funds were raised through respective local administrations), entitled *The Recasting of Science Between the Two World Wars*, with a deliberate multi-disciplinary approach and the contributions of acknowledged international specialists. Apart from the Varenna school of 1972, this meeting represents the official entrance of the group in the international academic environment as heralds of an original approach. The proceedings appeared some years later (Battimelli *et al.* 1984).

The path that led to the results cited in this section deserves to be more thoroughly reconstructed, we hope, with a collective engagement of many other colleagues.

**Conclusion**

In conclusion, under the stimulus of Kaiser's publication on the revival of the interest for the FQM in the US around the mid-1970s, we have reconstructed how an analogous revival in Italy dates back, more or less, to a decade prior.

Contrarily to the US, in Italy this interest had completely different motivations, mainly of a political nature. In fact, the young generation of the Italian physicists grew up in an environment following the student protests of 1968 and the subsequent working and popular struggles, in which the role of scientists in capitalistic society became an object of discussion and active engagement. In this complex situation, the interest of several young physicists was focused on the FQM, which was considered both as a way to deepen the ideological and material limitation of the kind of science developed in capitalistic society, and a search for possible alternatives. In parallel with this interest, the young generation of physicists felt the need to investigate the historical development of modern science, in the context of the social and economical conditions in which it developed.

In this paper we have offered a general review of this situation, and we have discussed the various fields of research and initiatives that it originated and the enduring results that it produced. One of these was precisely an active field of research on the FQM, which in many significant cases continued in the following decades. Regarding the field of history of science, in Italy there was a previous tradition in the Marxist "School of Geymonat", but the young physicists adopted a more radical point of view and quickly arrived to clash with this school. Consequently these young "outsiders" developed an original approach substantially still inspired by Marx, but based on his concept of historical materialism. They professionally developed this approach and carried out relevant initiatives, also at an international level, and contributed in renovating also the university curricula with the creation of specific courses on the history of physics.

---

[34] Roberto Livi and Stefano Ruffo performed the research for their thesis with Baracca.

**Acknowledgements**

We gratefully acknowledge the SIF for granting access to its archive and particularly Dr. Angela Oleandri, whose support has been irreplaceable in retrieving the data necessary to the completion of our work.

We are warmly grateful to Prof. David Kaiser and Prof. Olival Freire Jr. for their positive appreciation of our research, and for further information that have been useful in order to integrate our reconstruction.

We are moreover indebted to Jaxon Pope and Riccardo Centazzo for the precious revision of the English language of our manuscript.

This research did not receive any specific grant from funding agencies in the public, commercial, or not-for-profit sectors.

**References**

Amaldi, Edoardo. 1979. Gli anni della ricostruzione. *Giornale di Fisica*, pp. 186-22th5 (see http://www.phys.uniroma1.it/DipWeb/museo/collezione%20Fermi/fisicaAmaldi.htm: accessed 15[th] October, 2016).

Amaldi, Edoardo, Giovanni Battimelli and Giovanni Paoloni. 1998. *20[th] Century Physics: Essays and Recollections. A Selection of Historical Writings by Edoardo Amaldi*. World Scientific.

Augelli, Vincenzo, Augusto Garuccio, and Franco Selleri. 1976. La Mécanique Quantique et la Réalité. *Annales de la Fondation Louis de Broglie*. 1: 154-173.

Auletta, Gennaro (forwarded by Giorgio Parisi). 2000. *Foundations and Interpretation of Quantum Mechanics: In the Light of a Critical-Historical Analysis of the Problems and of a Synthesis of the Results.* World Scientific, Singapore.

Baracca, Angelo. 1970. *I problemi della misura e delle variabili nascoste nella teoria quantistica*, University of Florence, mimeographed lecture notes (December).

Baracca, Angelo. 1974. Il problema della causalità. general report at the Congress on the Foundations od Quantum Mechanics, Frascati, 4-6 June, unpublished.

Baracca, Angelo. 1980. *Manuale Critico di Meccanica Statistica*. CULC, Catania.

Baracca, Angelo and Silvio Bergia. 1971. Considerazioni critiche sulla metodologia della ricerca in fisica delle alte energie. Paper presented at the LVII National Congress of the Italian Society of Physics, L'Aquila, 26 October 1971, in *Bollettino SIF n. 87: 1-2*.

Baracca, Angelo and Silvio Bergia. 1972. Considerazioni critiche sulle scelte e sul metodo della ricerca attuale in fisica delle alte energie. Internal report, INFN/AE-72/4, Sezione di Bologna, 24 March 1972.

Baracca, Angelo and Silvio Bergia. 1976. *La Spirale delle Alte Energie: aspetti politici e logica di sviluppo della fisica delle particelle elementari*. Bompiani, Milan.

Baracca, Angelo, Silvio Bergia, Roberto Bigoni and Arinaldo Cecchini. 1974. Statistics of observations for proper and improper mixtures in quantum mechanics. *Rivista del Nuovo Cimento*. **4**: 169.

Baracca, Angelo, Silvio Bergia, Francesco Cannata, Stefano Ruffo, and Mirko Savoia. 1977. Is a Bell's type inequality for non dicotomic observable a good test of quantum mechanics?. *International Journal of Theoretical Physics*. **16**: 491.

Baracca, Angelo, Silvio Bergia and Flavio Del Santo. 2014. Le origini delle ricerche sui fondamenti della meccanica quantistica nel dopoguerra in Italia. Paper presented at the XXXIV Congress SISFA (Società Italiana di Storia della Fisica e dell'Astronomia). In *Atti del convegno SISFA, Firenze, 10-13 Settembre 2014*, edited by P. Tucci, in press.

Baracca, Angelo, Silvio Bergia, Roberto Livi, and Marta Restignoli. 1976. Reinterpretation and extension of Bell's inequality for multivalued variables. *International Journal of Theoretical Physics*. **15**: 473.

Baracca, Angelo, Silvio Bergia, and Amedeu Montoto. 1971. È possibile un'analisi consistente e indipendente dal modello dei processi di produzione a bassa energia?. Paper presented at the LVII National Congress of the Italian Society of Physics, L'Aquila, 26 October 1971, in *Bollettino SIF n 87: p. 1*.

Baracca, Angelo, Silvio Bergia, and Amedeu Montoto. 1972. Difficoltà nell'analisi dei processi di produzione a bassa energia e nella trattazione 'model independent' delle ampiezze a quasi due corpi", internal report INFN/AE-72/1, Sezione di Bologna, 24 January 1972.

Baracca, Angelo, Silvio Bergia and Marta Restignoli. 1974. On the comparison between quantum mechanics and local hidden variable theories: Bell's type inequality for multi-valued observables. *Proceedings of the International Conference on Few Body Problems in Nuclear and Particle Physics*, Quèbec, 27-31, August, p. 67.

Baracca, Angelo, David J. Bohm, Basil J. Hiley and Allan. E. G. Stuart. 1975. On some notions concerning locality and non locality in the quantum theory. *Il Nuovo Cimento* **28B**: 435.

Baracca, Angelo, Silvio Bergia, Francesco Cannata, Stefano Ruffo and Mirko Savoia. 1977. Is a Bell-Type Inequality for Nondicotomic Observables a Good Test of Quantum Mechanics?. *International Journal of Theoretical Physics* **16**: 491.

Baracca Angelo, Alfonso Cornia, Roberto Livi, and Stefano Ruffo. 1978a. Quantum mechanics, first kind states and local hidden variables: three experimentally distinguishable situations. *Il Nuovo Cimento B* **43**: 65.

Baracca, Angelo, Alfonso Cornia, Antonio Lunardini and Ruffo Stefano. 1978b. On the tests of quantum mechanics: how many different theoretical descriptions?. *Lettere al Nuovo Cimento* **22**: 281.

Baracca Angelo, Saverio Craparo, Roberto Livi and Stefano Ruffo. 2016. The Role of Physics Students from the University of Florence (later Professors) in the Early Italian Anti-nuclear Movements (1975-1987). In *Nuclear Italy: An International History of Italian Nuclear Policies during the Cold War,* Proceedings of the Conference on Italy's Nuclear Experience in an International and Comparative Perspective, University of Trieste, November 13-15, 2014 edited by Elisabetta Bini and Igor Londero. EUT, Trieste, in press.

Baracca, Angelo and Carlo De Marzo. 1972. Appunti sulla storia della fisica. *Giornale di Fisica* **13**: 67.

Baracca, Angelo and Roberto Livi. 1976. *Natura e Storia, Fisica e Sviluppo del Capitalismo nell'Ottocento*. D'Anna, Firenze.

Baracca, Angelo and Riccardo Rigatti. 1974a. Aspetti dell'interazione tra scienza e tecnica durante la rivoluzione industriale inglese del XVIII° sec. in Inghilterra, 1ª Parte: La nascita dei concetti di lavoro ed energia. *Giornale di Fisica* **15**: 144.

Baracca, Angelo and Riccardo Rigatti. 1974b. Aspetti dell'interazione tra scienza e tecnica durante la rivoluzione industriale inglese del XVIII° sec. in Inghilterra, 2ª Parte: Sviluppo della macchina a vapore, *Giornale di Fisica* **15**: 206.

Baracca, Angelo, and Arcangelo Rossi. 1976. *Marxismo e Scienze Naturali: Per una Storia Integrale delle Scienze*. De Donato, Bari.

Baracca, Angelo, Stefano Ruffo and Arturo Russo. 1979. *Scienza e Industria, 1848-1915*. Laterza, Bari.

Battimelli, Giovanni, Michelangelo De Maria and Arcangelo Rossi. 1984. *La Ristrutturazione delle Scienze tra la Due Guerre Mondiali* (2 Volumes). La Goliardica Editrice Universitaria, Roma.

Bellone, Enrico, Ludovico Geymonat, Giulio Giorello, Silvano Tagliagambe. 1974. *Attualità del Materialismo Dialettico.* Editori Riuniti, Roma.

Bergia, Silvio and Francesco Cannata. 1982. Higher-Order Tensors and Tests of Quantum Mechanics. *Foundations of Physics* **12**: 843.

Bergia, Silvio, Francesco Cannata, Alfonso Cornia and Roberto Livi. 1980. On the Actual Measurability of the Density Matrix of a Decaying System by Means of Measurements on the Decay Products. *Foundations of Physics* **10**: 723.

Bergia, Silvio, Francesco Cannata and Vittorio Monzoni. 1985. Explicit Examples of Theories Satisfying Bell's Inequality: Do They Miss Their Goal Prior to Contradicting Experiments?. *Foundations of Physics* **15**: 145.

Bergia, Silvio, Francesco Cannata and Antonello Pasini. 1988. Space Time Fluctuations and Stochastic Mechanics: Problems and perspectives, in L. Kostro, A. Posiewnik, J. Pycacz, M. Zukowski (Eds.), *Problems in Quantum Physics,* Gdansk '87, World Scientific, pp. 403-421.

Bergia, Silvio, Francesco Cannata and Antonello Pasini. 1989. On the possibility of interpreting quantum mechanics in terms of stochastic metric fluctuations. *Physics Letters* **137 A**: 21.

Bergia, Silvio, Francesco Cannata, Stefano Ruffo and Mirko Savoia. 1979. Group theoretical interpretation of von Neumann's theorem on composite systems. *American Journal of Physics* **47**: 548.

Bergia, Silvio and Paola Fantazzini. 1981. La descrizione dei fenomeni meccanici in termini energetici nell'opera di John Smeaton; Parte prima: Dalle ruote ad acqua ai teoremi dell'impulso e delle forze vive. *Giornale di Fisica* Vol. XXII **4**: 295-310.

Bergia, Silvio and Paola Fantazzini. 1982. La descrizione dei fenomeni meccanici in termini energetici nell'opera di John Smeaton; Parte seconda: Un modello per l'urto anelastico ed il suo controllo sperimentale. *Giornale di Fisica* Vol.XXIII, **1**:. 59-73.

Bergia, Silvio, Carlo Ferrario and Vittorio Monzoni. 1985a. Side Paths in the History of Physics: The Idea of Light Molecule from Ishiwara to de Broglie. *Rivista di Storia della Scienza* Vol. 2, n. 1, **71**: 71-97.

Bergia, Silvio, Carlo Ferrario and Vittorio Monzoni. 1985b. Planck's heritage and the Bose statistics. *Annales de la Fondation Louis de Broglie* **10**: 161.

Bergia, Silvio, Paolo Lugli and Nadia Zamboni. 1979. Zero-point energy, Planck's law and the prehistory of stochastic electrodynamics. Part 1: Einstein and Hopf's paper of 1910. *Annales de la Fondation Louis de Broglie* **4**: 295.

Bergia, Silvio, Paolo Lugli and Nadia Zamboni. 1980. Zero-point energy, Planck's law and the prehistory of stochastic electrodynamics. Part 2: Einstein and Stern's paper of 1913. *Annales de la Fondation Louis de Broglie* **5**: 39.

Boccaletti, Dino and Franco Selleri. 1961. Some remarks on double photoproduction. *Il Nuovo Cimento (1955-1965)* **22.5**: 1099-1103.

Bonsignori, Francesco and Franco Selleri. 1960. Pion cloud effects in pion production experiments. *Il Nuovo Cimento* **15.3**: 465-478.

Caldirola, Piero. 1961. *Quantistica, Meccanica*, entry in the *Enciclopedia Italiana*, III$^{rd}$ Appendix (XXVIII, p. 592; App. II, 11, p. 634), http://www.treccani.it/enciclopedia/meccanica-quantistica_res-9300d5ef-87e8-11dc-8e9d-0016357eee51_(Enciclopedia-Italiana)/, accessed 15$^{th}$ Oactober, 2016.

Capasso, Vincenzo, Donato Fortunato and Franco Selleri. 1970. von Neumann's theorem and hidden variables models. *Rivista del Nuovo Cimento,* **2**: 149-199.

Capasso, Vincenzo, Dino Fortunato and Franco Selleri. 1973. Sensitive observables in quantum mechanics. *International Journal of Theoretical Physics* **7**: 319-326.

Ciccotti, Giovanni, Marcello Cini, Michelangelo De Maria, Giovanni Jona-Lasinio, Elisabetta Donini and Narducci Dario. 1976. *L'Ape e l'Architetto: Paradigmi Scientifici e Materialismo Storico*. Feltrinelli, Milan.

Ciliberto, Sergio, Saverio Craparo, Giovanni Del Fante, Roberto Livi, Marco Lugli, Marco Pettini, Antonio Politi, Andrea Raspini, and Lorenzo Vallerini. 1977. *I Nucleodollari.* CP Editrice, Florence.

Cini, Marcello. 1969. Il Satellite della Luna. *il manifesto* (rivista), September.

Clauser, John F. 1992. Early history of Bell's theory and experiment. In T. D. Black et al (eds), Foundations of Quantum Mechanics, (pp. 168-74). Singapore: World Scientific.

Cufaro Petroni, Nicola and Jean Pierre Vigier. 1979. On Two Conflicting Physical Interpretations of the Breaking of Restricted Relativistic Einsteinian Causality by Quantum Mechanics. *Lettere al Nuovo Cimento* 25: 151.

Daneri, Adriana., Loinger Angelo, and Prosperi Giovanni Maria, 1962, Quantum theory of measurement and egodicity conditions, *Nuclear Physics*, 33, 297 *(reprinted in John Archibald Wheeler and Woj*ciech Hubert Zurek, eds., *Quantum Theory and Measurement*, Princeton, Princeton University Press, 1983, pp. 657–679).

D'Espagnat, Bernard. 1965. *Conceptions de la Physique Contemporaine: les Interprétations de la Mécanique Quantique et de la Mesure*. Vol. 1320. Hermann.

D'Espagnat, Bernard. 1966. Two remarks on the Theory of Measurement. *Supplemento al Nuovo Cimento* **1**: 828–838.

D'Espagnat, Bernard. 1971. *Proceedings of the International School of Physics "Enrico Fermi", Foundations of Quantum Mechanics 1970,* edited by SIF, Vol. 49. Academic Press.

Dieks, Dennis. 1982. Communication by EPR devices. *Physics Letters A* **92 (6)**: 271–272.

Donini, Elisabetta, Arcangelo Rossi and Tito Tonietti. 1977. *Matematica e Fisica: Struttura e Ideologia*. De Donato, Bari.

Fabri, Elio. 1970. Lettera del Prof. E. Fabri. *Bollettino SIF n. 77, pp. 8-12*.

Falciglia, Filippo, Augusto Garuccio and Lorenzo Pappalardo. 1982. Rapisarda's experiment: on the four-coincidence equipment 'FOCA-2', a test for nonlocality propagation. *Lettere al Nuovo Cimento* **34**: 1-4.

Faraci, Giuseppe, Diego Gutkowski, Salvatore Notarrigo and Agata Pennisi. 1974. An experimental test of the EPR paradox. *Lettere al Nuovo Cimento* **9.15**: 607-611.

Ferrari, Ezio and Franco Selleri. 1963. An approach to the theory of single pion production in nucleon-nucleon collisions. *Il Nuovo Cimento* **27.6**: 1450-1483.

Fortunato, Dino, Augusto Garuccio, and Franco Selleri. 1977. Observable consequences from second-type state vectors of quantum mechanics. *International Journal of Theoretical Physics* **16**: 1-6.

Fortunato, Dino and Franco Selleri. 1976. Sensitive observables in infinite-dimensional Hilbert spaces. *International Journal of Theoretical Physics* **15**: 333-338.

Freedman, Stuart J. and John F. Clauser. 1972. Experimental test of local hidden-variable theories. *Physical Review Letters* **28.14**: 938.

Fonda, Luciano, Gian Carlo Ghirardi, Alberto Rimini and Tullio Weber. 1973. On the Quantum Foundations of the Exponential Decay Law. *Nuovo Cimento* **15A**: 689; Ivi. **18A**: 805.

Freire, Olival Jr. 2003. A Story Without an Ending: The Quantum Physics Controversy 1950–1970. *Science & Education* **12**: 573–586.

Freire, Olival Jr. 2004. Popper, probabilidade e mecânica quântica. *Episteme* **18**:103-127

Freire, Olival Jr. 2006. Philosophy enters the optics laboratory: Bell's theorem and its first experimental tests (1965–1982), *Studies in History and Philosophy of Science, Part B: Studies in History and Philosophy of Modern Physics* **37.4**: 577-616.

Freire, Olival Jr.. June 24, 2003. Interview to Franco Selleri. Sessions I and II (https://www.aip.org/taxonomy/term/2126, accessed 15[th] October, 2016).

Freire, Olival Jr. 2014. *The Quantum Dissidents: Rebuilding the Foundations of Quantum Mechanics (1950-1990)*. Springer, Berlin.

Freire, Olival Jr. and Osvaldo Jr. Pessoa. 2015. Bell's theorem without inequalities: on the inception and scope of the GHZ theorem. *International Journal of Quantum Foundations*, online forum, http://www.ijqf.org/archives/1767, accessed 15[th] October, 2016.

Gagliasso, Elena, Mattia Della Rocca, Rosanne Memoli. 2015. *Per una scienza critica. Marcello Cini e il presente: filosofia, storia e politiche della ricerca*. Edizioni ETS, Pisa. http://www.edizioniets.it/Priv_File_Libro/2679.pdf, accessed 15[th] October, 2016.

Garuccio, Augusto, Karl R. Popper, and Jean-Pierre Vigier. 1981. Possible direct physical detection of de Broglie waves. *Physics Letters A* **86.8**: 397-400.

Garuccio, Augusto and Vittorio Rapisarda. 1981. Bell's inequalities and the four-coincidence experiment. *Nuovo Cimento* **65A**: 289.

Garuccio, Augusto, Vittorio Rapisarda and Jean-Pierre Vigier. 1982. New experimental set-up for the detection of de Broglie waves. *Physics Letters* **90**: 17-19.

Garuccio, Augusto, Giancarlo Scalera and Franco Selleri. 1977. On local causality and the quantum-mechanical state vector. *Lettere al Nuovo Cimento* **18**: 26-28.

Garuccio, Augusto and Franco Selleri. 1976. Nonlocal interactions and Bell's inequality. *Nuovo Cimento* **36B**: 176-85.

Garuccio, Augusto, and Jean Pierre Vigier. 1980. Possible experimental test of the causal stochastic interpretation of quantum mechanics: Physical reality of de Broglie waves. *Foundations of Physics* 10.9-10: 797-801.

Ghirardi, Gian Carlo, Claudio Omero, Alberto Rimini and Tullio Weber. 1978. The Stochastic Interpretation of Quantum Mechanics: a Critical Review. *Rivista del Nuovo Cimento* 1.3: 1-34.

Ghirardi, Gian Carlo, Alberto Rimini and Tullio Weber. 1976. Implications of the Bohm-Aharonov Hypothesis. *Il Nuovo Cimento* 31B: 177.

Ghirardi, Gian Carlo, Alberto Rimini and Tullio Weber. 1977. Some Simple Remarks about Quantum Nonseparability for Systems Composed of Identical Constituents. *Il Nuovo Cimento* 39B: 130.

Ghirardi, Gian Carlo, Alberto Rimini and Tullio Weber. 1984. A Model for a Unified Quantum Description of Macroscopic and Microscopic Systems Quantum Probability and Applications. In *Quantum Probability and Applications II: Proceedings of a Workshop held in Heidelberg, West Germany, October 1-5, 1984*, edited by L. Accardi et al., *Quantum Probability and Applications II*, Springer, Berlin, pp. 223-232.

Ghirardi, Gian Carlo and Tullio Weber. 1979. On Some Recent Suggestions of Superluminal Communication through the Collapse of the Wave Function. *Lettere Nuovo Cimento*. **26**: 599.

Ghirardi, Gian Carlo, Alberto Rimini and Tullio Weber. 1986. Unified dynamics for microscopic and macroscopic systems. *Physical Review D* **34.2**: 470.

Guidoni, Paolo. 1970. Risonanze bosoniche 1970. Invited paper at the LVI Congress SIF, Venice 1970, *Bollettino* SIF n. 79, October 15, p. 32.

Herbert, Nick. 1982. FLASH—A superluminal communicator based upon a new kind of quantum measurement. *Foundations of Physics* **12(12):** 1171–1179.

Kaiser, David. 2011. *How the hippies saved physics: science, counterculture, and the quantum revival*. WW Norton & Company, New York.

Krips, Henry. 2007. *Measurement in Quantum Theory*, Stanford Encyclpedia of Philosophy (*First published October 12, 1999; substantive revision August 22, 2007),* http://plato.stanford.edu/archives/sum2016/entries/qt-measurement/, accessed 15[th] October, 2016.

Livi, Roberto. 1977. New Tests of Quantum Mechanics for Multivalued Observables. *Lettere Nuovo Cimento* **19**: 272.

Livi, Roberto. 1978. An Experimental test of Quantum Mechanics in Molecular Predissociation. *Il Nuovo Cimento* B **48**: 272.

Maltese, Guido. 2010. *Il Papa e l'Inquisitore. Enrico Fermi, Ettore Majorana, via Panisperna. Zanichelli, Bologna.*

Mara, Luigi. 2010. *Scienza, salute e ambiente. L'esperienza di Giulio Maccacaro e di Medicina Democratica, Pristem/Storia (Bocconi University), 27-28, Il '68 e la Scienza in Italia* (edited by Angelo Guerraggio), p. 49-72.

Nutricati, Pompilio. 1998. *Oltre I Paradossi della Fisica Moderna: I Fisici Italiani per il Rinnovamento di Teoria Quantistica e

*Relatività. Dedalo, Bari.*

Popper, Karl R. 1982. *Quantum Theory and the Schism in Physics (The Postscript to The logic of scientific discovery, Vol.III*. Rowman And Littlefield, Totowa, New Jersey.

Rossi, Arcangelo. 2010. *L'esperienza di "Testi e Contesti", Pristem/Storia (Bocconi University), 27-28, Il '68 e la Scienza in Italia* (edited by Angelo Guerraggio), p. 87-96.

Selleri, Franco. 1961. Role of the Peripheral Interaction in Proton-Proton Collisions. *Physical Review Letters* **6**: 64-66.

Selleri, Franco. 1969a. On the wave function in quantum mechanics. *Lettere al Nuovo Cimento Vol 1* **17**: *908-910.*

Selleri, Franco. 1969b. Quantum Theory and Hidden Variables. Unpublished results, lectures held in Frascati (June-July), LNF – 69/75 CNEM-Laboratori Nazionali di Frascati

Selleri, Franco. 1970. La piramide azteca della fisica teorica delle particelle elementary. *Bollettino SIF,* **75**: 13-17.

Selleri, Franco. 1972. A stronger form of Bell's inequality, On local causality and the quantum-mechanical state vector. *Lettere al Nuovo Cimento* **3**: 581-82.

Selleri, Franco. 1980. Einstein locality and the quantum mechanical long-distance effects, in A. **Blaquiere**, F. **Fer and** A. **Marzollo** (Eds.). Springer, Wien. 393-412.

Selleri, Franco. 1986, *El Debate de la Teorìa Cuántica*. Alianza Editorial, Madrid.

Selleri, Franco and Jean Pierre Vigier. 1980. Unacceptability of the Pauli-Jordan propagator in physical applications of quantum theory. *Lettere al Nuovo Cimento* **29**: *pp. 7-9.*

Shields, William M. 2012. A Historical Survey of Sir Karl Popper's Contribution to Quantum Mechanics. *Quanta* **1.1**: 1-12.

SIF, Bollettino, 1968, n. 62, 21 October 1968, p. 7.

SIF. Bollettino, 1970, n. 74, 13 April 1970, pp. 6-29.

SIF, Bollettino, 1970, n. 76, 8 June 1970, pp. 7-8.

SIF, Bollettino, 1971, n. 84, 15 July 1971, pp. 4-12.

SIF (Ed.), 1971, *La Scienza nella Società Capitalistica*, Società Italiana di Fisica (Eds.), Bari, De Donato.

SIF, SC, 1968, Minutes of the Steering Committee, 16 November 1968, SIF Archives, Bologna.

SIF, SC, 1969a, Minutes of the Steering Committee, 15 March 1969, SIF Archives, Bologna.

SIF, SC, 1969b, Minutes of the Steering Committee, 19 April 1969, SIF Archives, Bologna.

SIF, SC, 1969c, Minutes of the Steering Committee, 11 October 1969, SIF Archives, Bologna.

SIF, SC, 1970a, Minutes of the Steering Committee, 7 February 1970, SIF Archives, Bologna.

SIF, SC, 1970b, Minutes of the Steering Committee, 2 March 1970, SIF Archives, Bologna.

SIF, SC, 1970c, Minutes of the Steering Committee, 26 March 1970, SIF Archives, Bologna.

SIF, SC, 1970d, Minutes of the Steering Committee, 18 April 1970, SIF Archives, Bologna.

SIF, SC, 1970e, Minutes of the Steering Committee, 31 May 1970, SIF Archives, Bologna.

SIF, SC, 1970f, Minutes of the Steering Committee, 18 July 1970, SIF Archives, Bologna.

SIF, SC, 1970e, Minutes of the Steering Committee, 9 September 1970, SIF Archives, Bologna.

SIF, SC, 1971a, Minutes of the Steering Committee, 25 March 1971, SIF Archives, Bologna.

SIF, SC, 1971b, Minutes of the Steering Committee, 8 May 1971, SIF Archives, Bologna.

SIF, SC, 1971c, Minutes of the Steering Committee, 25 May 1971, SIF Archives, Bologna.

Tagliagambe, Silvano (editor). 1972. *L'Interpretazione Materialistica della Meccanica Quantistica. Fisica e Filosofia in URSS.* Feltrinelli, Milan.

Tarozzi, Gino, and Alwyn van der Merwe. 1985. *Open questions in quantum physics*. Reidel, Dordrecht.

VV. AA., 1970, "Notes on the connection between science and society". Unpublished results. Mimeographed document produced by a group of participants to the Summer Course on the Foundation of Quantum Mechanics, International School of Physics, Varenna (Archive of the Italian Physical Society, Bologna, box of documents about Varenna's school 1970 on FQM).

Weiner, Charles. *1977. Proceedings of the International School of Physics "Enrico Fermi", History of Twentieth Century Physics 1972*, edited by SIF, Academic Press.

Wootters, William K., Zurek Wojciech H. 1982. A single quantum cannot be cloned. *Nature* **299.5886**: 802-803.